\documentclass[runningheads]{llncs}
\usepackage[T1]{fontenc}
\usepackage{graphicx}

\usepackage{hyperref}
\begin{document}

\title{Let's Get It Started: Fostering~the Discoverability~of~New Releases on Deezer}

\titlerunning{Let's Get It Started}
%

\author{Léa Briand$^{1,*}$, Théo Bontempelli$^{1}$, Walid Bendada$^{1,2}$, \\ Mathieu Morlon$^{1}$,  François Rigaud$^{1}$, Benjamin Chapus$^{1}$, \\Thomas Bouabça$^{1}$,  \and Guillaume Salha-Galvan$^{1}$}
\authorrunning{L. Briand, et al.}

%
\institute{$^{1}$Deezer Research, Paris, France \\
$^{2}$LAMSADE, Université Paris Dauphine, PSL, Paris, France \\
$^{*}$Corresponding author at:
\email{research@deezer.com}
}
\maketitle              
\begin{abstract}
This paper presents our recent initiatives to foster the discoverability of new releases on the music streaming service Deezer. 
After introducing our search and recommendation features dedicated to new releases, we outline our shift from editorial to personalized release suggestions using cold start embeddings and contextual bandits. Backed by online experiments, we discuss the advantages of this shift in terms of recommendation quality and exposure of~new~releases on the service.
\\ $ $ \\ \textbf{Note:} This short article presents a work that has been accepted for oral presentation as an ``Industry Talk'' at the 46th European Conference on Information Retrieval (ECIR 2024). Resources related to this talk will be available on: \href{https://github.com/deezer/new-releases-ecir2024}{\texttt{https://github.com/deezer/new-releases-ecir2024}}.

\keywords{New Releases \and Music Recommendation \and Music Discovery.}
\end{abstract}
\section{Introduction}

Music artists release hundreds of thousands of new albums every week on the music streaming service Deezer~\cite{deezerwebsite}.
The prompt integration of this content, along with its swift discoverability through recommender systems and search engines, holds significant importance. For Deezer, it ensures that users have immediate access to the latest music of their favorite artists while also easily coming upon new ones they might like, which is known to improve the user experience~\cite{li2023recent,schedl2018current}. The proper
exposure of new releases also benefits artists by amplifying their visibility, which
can contribute to their success, boost their revenues, and foster the emergence
of new talents~\cite{aguiar2018platforms,ferraro2021fair}.
Nonetheless, displaying the right releases to the right users remains challenging due to the limited prior information on this fresh content, especially for new artists unknown from~the~service~\cite{celma2010long,schedl2018current}.

In this paper, we present our recent efforts to better showcase new releases on Deezer, both in terms of recommendation performance and number of new releases exposed to users. Specifically, we first focus on the product context. We describe in Section~2 our search and recommendation features dedicated to new releases, along with their objectives and differences. 
We also dive into our historical semi-personalized solution for new release recommendation, based on editorial pre-selections, and its limitations in terms of catalog coverage and adaptability.
Then, in Section 3, we detail the fully personalized systems we deployed in 2023 to overcome these limitations, involving a cold start neural embedding model along with contextual bandits~\cite{briand2021semi,chapelle2011empirical}. We provide an online evaluation of our approach on Deezer's ``New releases for you'' carousel of recommended new albums.
Finally, we conclude in Section 4 by discussing our ongoing efforts to improve our models and how we present new releases on~the~Deezer~homepage.

\begin{figure}[t!]
  \centering
  \includegraphics[width=1\linewidth]{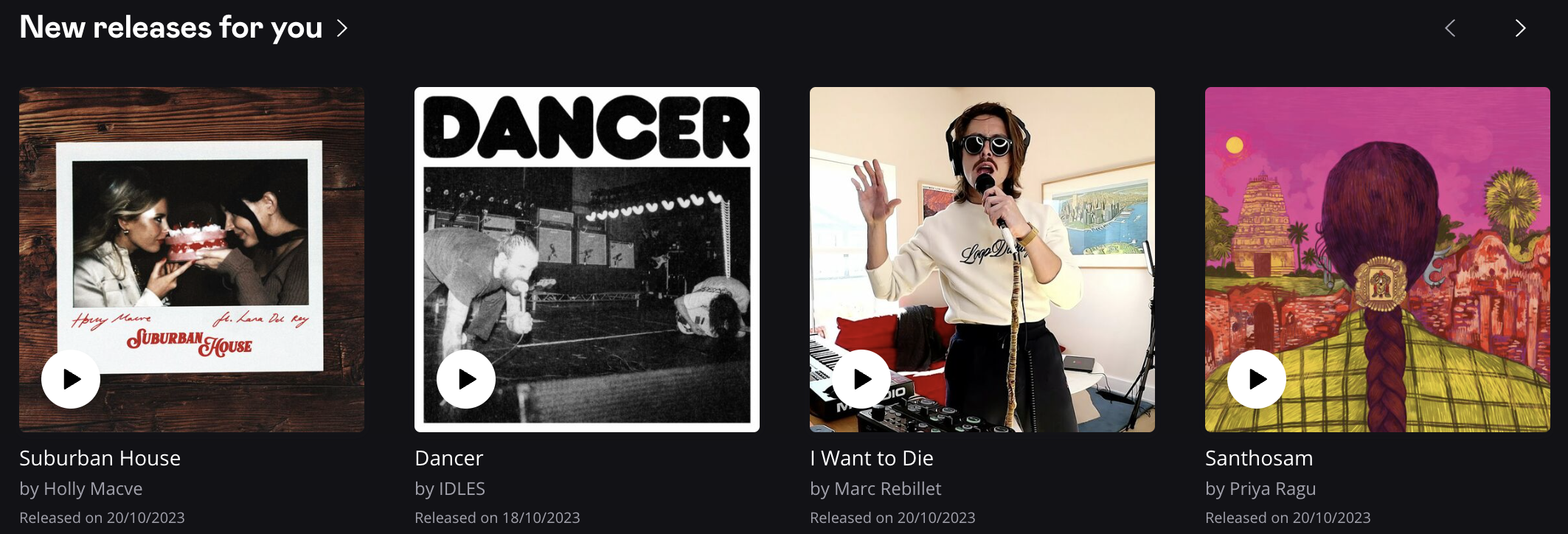}
  \caption{Interface of the ``New releases for you'' carousel on the website version of Deezer, recommending a personalized shortlist of recently released albums to each~user.}
\label{figure-nrfy}
\end{figure}

\section{Listening to New Releases on Deezer}

\label{newreleasesfeatures}

Hundreds of thousands of albums and music tracks are released on Deezer every week.
Users can instantly access this content using our \textit{search engine}, based on an Elasticsearch index plugged into a Kafka topic~\cite{elasticsearch,kafka} refreshed~in~real-time. 
They also receive \textit{notifications} informing them about updates from their favorite artists, which we refer to as \textit{unmissable} releases. In addition, on the Deezer homepage, we propose ``\textit{Friday cards}'', i.e., weekly playlists mixing unmissable~tracks.

All these features help users retrieve releases from artists they already know. However, they hardly permit \textit{discovering} new~music from emergent or unfamiliar artists \cite{celma2010long}.
For this distinct purpose, the Deezer homepage relies on two album carousels\footnote{Our carousels display swipeable ranked lists of 12 albums, followed by a ``View all'' button leading to a page displaying up to 100 recommended albums.}~\cite{bendada2020carousel}. The first one, known as ``\textit{Fresh~picks~of~the~week}'', presents the biggest releases according to professional editors~from~Deezer.
The second one, called ``\textit{New~releases~for~you}'' and illustrated in Figure~\ref{figure-nrfy}, aims to provide personalized recommendations. Up to early 2023, it first showcased unmissable albums for each user, followed by new albums from their favorite music genres. 
Our data scientists were identifying each user's favorite genres, but the lists of recommended albums for each genre were manually pre-selected~by~Deezer~editors.

This historical method for discovering new releases had limitations that motivated the work presented in this paper. To begin with, ``New~releases~for~you'' carousels were only partially personalized. All fans from a genre used to receive the same (non-unmissable) recommendations, regardless of their individual subgenre preferences.
Moreover, editorial lists were only updated once a week, on Fridays. Carousels remained static for seven days, making the feature less engaging. New releases added during the week went unrecommended for days, and user interactions were not factored in for improved recommendations.
Last but not least, only the small minority of albums handpicked by editors were recommended. Most new releases never got the chance to even be exposed to users.

\section{Improving the Discoverability of New Releases}

This section outlines our recent initiatives to overcome these limitations. For the sake of clarity, we focus our presentation on the discovery of \textit{new albums}.
\subsection{Cold Start Embedding Representations of New Releases}

\label{coldstartsection}

Most of our personalized recommender systems~\cite{bendada2023track,bendada2020carousel,bontempelli2022flow,briand2021semi,salha2021cold} leverage latent models for collaborative filtering (CF)~\cite{bokde2015matrix,koren2015advances}. By analyzing usage data on Deezer, they learn low-dimensional \textit{embedding} vector representations of users and musical items including albums, in a vector space where proximity reflects~preferences.
Then, they offer recommendations based on embedding similarity metrics~\cite{bontempelli2022flow,briand2021semi}. 
Unfortunately, such CF models often struggle to incorporate new items lacking usage data \cite{mu2018survey,schedl2018current,wang2018billion}. This issue, known as the \textit{cold start} problem, explains why Deezer historically favored an editorial approach for suggesting~new~releases.

In 2023, we developed CF-Cold-Start. For any existing CF embedding space, CF-Cold-Start predicts future embedding vectors of items such as new albums \textit{as early as their release date}, without requiring extensive usage~data. CF-Cold-Start draws inspiration from advances in cold start research \cite{covington2016deep,lee2019melu,wang2018billion}, including a Deezer system for \textit{user} cold start~\cite{briand2021semi}. 
 It consists in a 3-layer neural network~\cite{goodfellow2016deep}, predicting ``ground truth'' embedding vectors that would be computed by the CF model using a week of usage data.
In the case of a new album, the input layer processes various album metadata, including information on labels and artists, as well as usage data (streams, likes...) at the release date when available. The output layer returns a \textit{predicted embedding vector} for this~album. We train CF-Cold-Start in a supervised manner using musical items for which both input data and ground truth embedding vectors 
are available, by minimizing the mean squared error between predicted and ground truth vectors using gradient~descent~\cite{goodfellow2016deep}.

\begin{figure}[t]
  \centering
  \includegraphics[width=1\linewidth]{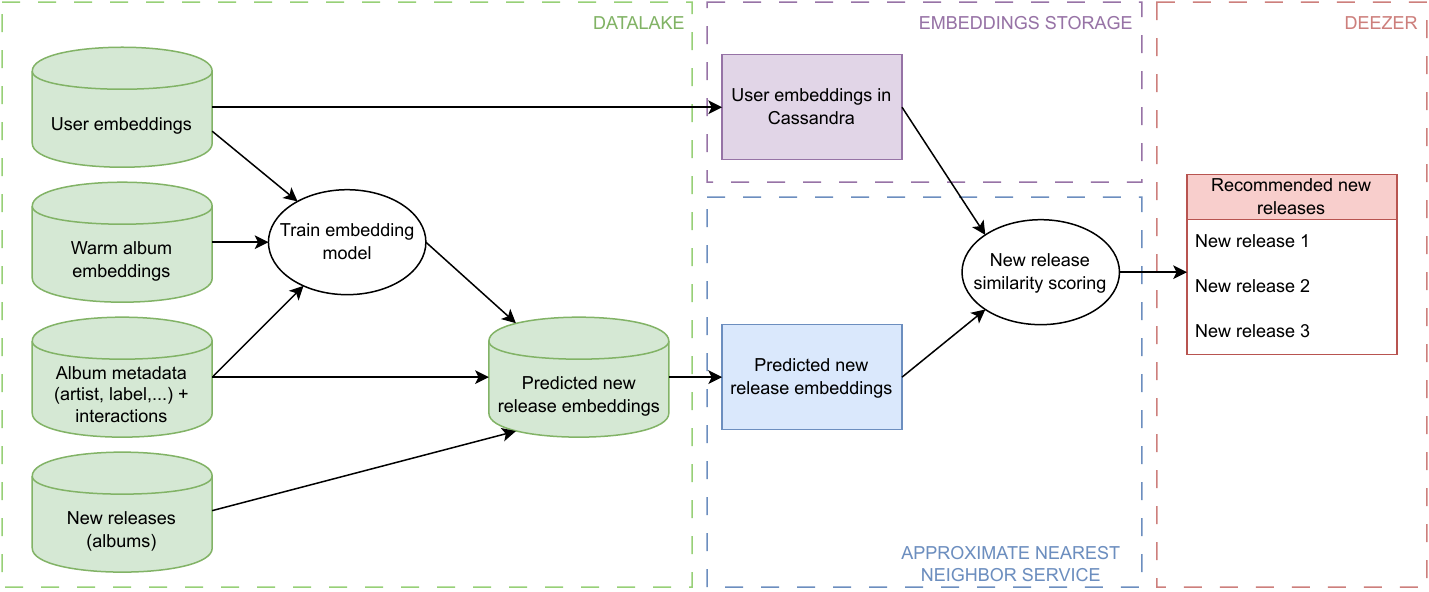}
  \caption{Overview of the CF-Cold-Start new album recommendation framework from Section~\ref{coldstartsection}, available for online requests in our production environment on Deezer. }
\label{figure-coldstart}
\end{figure}

We deployed CF-Cold-Start on Deezer in our ``New releases for you''~feature. While carousels still begin with unmissable new albums, CF-Cold-Start replaces editorial pre-selections with personalized~recommendations, consisting in ranked lists of albums whose predicted embedding vectors are the most similar to each user\footnote{Precisely, this feature uses dot product similarities and the SVD embeddings~from~\cite{bendada2023track}.} in the CF embedding space.
We update embedding vector predictions every~four~hours through a forward pass in CF-Cold-Start, to refine recommendations as we collect more usage data on new albums. 
We use approximate nearest neighbors methods via a Golang application incorporating the Faiss~library~\cite{johnson2019billion}. 
Figure~\ref{figure-coldstart} provides an illustrated summary of our CF-Cold-Start new album recommendation pipeline in production.



\subsection{Carousel Personalization with Contextual Bandits}

\label{bandit}

CF-Cold-Start offers fully personalized recommendations from the entire set of new albums, with refreshes every four hours instead of once~a~week. Nonetheless, it solely \textit{exploits} the most tailored albums based on usage data. In this work, we also wanted to occasionally deviate from them, to \textit{explore} less popular but promising albums which could be overlooked by relying on the existing usage.

To this end, we developed TS-CF-Cold-Start, a CF-Cold-Start~variant with a \textit{multi-arm bandit} component for adaptive ``New releases for you'' album ranking. Bandit algorithms effectively handle exploration-exploitation face-offs~\cite{chapelle2011empirical,li2010contextual,mcinerney2018explore}. Here, we use the \textit{Thompson Sampling}~(TS) \cite{thompson1933likelihood} extension for contextual bandits \cite{chapelle2011empirical}. Our \textit{arms} are new albums, recommended by ranked batches at each bandit \textit{round}, i.e., each carousel display on Deezer. The set of arms evolves over time, as new albums enter daily and exit the new release set after seven days. 

\begin{figure}[t]
  \centering
  \includegraphics[width=1\linewidth]{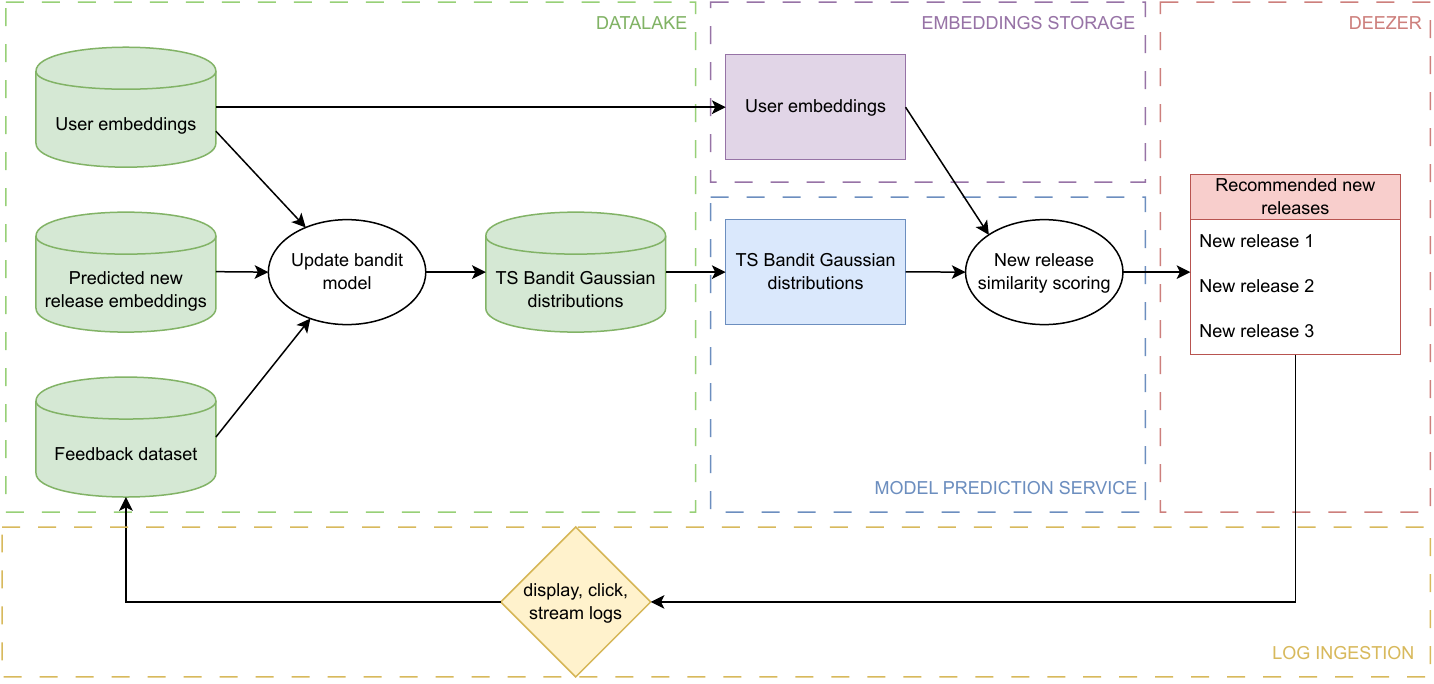}
  \caption{Overview of the TS-CF-Cold-Start new album recommendation framework from Section~\ref{bandit}, available for online requests in our production environment on Deezer. }
\label{figure-bandit}
\end{figure}

Album embedding vectors predicted by CF-Cold-Start act as the \textit{prior expected arm representations}~\cite{chapelle2011empirical} used for arm selection by the bandit. They are updated every four hours using TS with Gaussian distributions. Each album's mean and variance are exported in an ONNX model \cite{onnx} in our service for inference and sampling.
Updates are based on user click \textit{rewards} and a \textit{cascade-based} scheme~\cite{bendada2020carousel} accounting for the album position in the carousel. We note that, while carousels still start with fixed unmissable releases, the stochastic nature of TS-based exploration-exploitation permits recommending different non-unmissable albums with each homepage refresh, even between~two~arm~updates.
Figure~\ref{figure-bandit} summarizes our TS-CF-Cold-Start pipeline in production.

\subsection{Online Experimental Evaluation on Deezer}

We conducted large-scale A/B tests on Deezer between March~and~May~2023. 
On average, replacing editorial selections with CF-Cold-Start recommendations in ``New releases for you'' not only improved display-to-click rates by 6\%, but also multiplied by $\times$3 and $\times$1.5 the weekly numbers of new albums displayed and clicked in carousels, respectively. Hence, CF-Cold-Start boosted their exposure.

Using TS-CF-Cold-Start added dynamism to carousels by sampling albums with each app refresh. However, its performance was on par with CF-Cold-Start in both clicks and exposure. 
This mixed result contrasts with a previous study~\cite{bendada2020carousel} where bandits improved our \textit{editorial playlist} carousels.
We attribute our lower performance to two factors. Firstly, unmissable albums restrict bandit actions to less visible carousel slots, unlike in playlist carousels~\cite{bendada2020carousel}. Secondly, our set of new albums constantly evolves, potentially hindering model~convergence. 

Note: our experiments will be further illustrated and discussed in the ECIR ``Industry~Talk'' associated with this article. Resources related to this talk will be available on: \href{https://github.com/deezer/new-releases-ecir2024}{\texttt{https://github.com/deezer/new-releases-ecir2024}}. 

\section{Conclusion and Future Work}

In conclusion, both CF-Cold-Start and TS-CF-Cold-Start enhanced our ``New releases for you'' recommendations and the exposure of new albums, all of this with more dynamic updates. These initiatives foster the overall discoverability of new releases on Deezer. They also open up interesting avenues for future work. We believe our problem is conducive to the study of bandits explicitly accounting for \textit{equity of exposure}, for improved fairness in release suggestions~\cite{jeunen2021top}.
From a product perspective, our mixed results with TS-CF-Cold-Start also prompt us to separate unmissable albums from discoveries into two carousels, with the latter fully controlled by bandits. Lastly, we will use CS-Cold-Start to better incorporate releases in other Deezer features, including our ``Flow''  personalized~radio~\cite{bontempelli2022flow}.

\section*{Speaker and Company}

\textbf{Léa Briand} is a Senior Data Scientist at \textbf{Deezer}, a French music streaming service created in 2007 and with over 16~million active users in 180 countries. In the Recommendation team, she develops large-scale machine learning systems to improve music recommendation on this service.

\bibliographystyle{splncs04}
\bibliography{references}

\end{document}